\documentclass[aps,prb,reprint,showpacs,superscriptaddress,longbibliography]{revtex4-2}
\usepackage{mathtools,amssymb,graphicx,units}

\usepackage[plainpages=false,pdfpagelabels,colorlinks=true,linkcolor=red,urlcolor=blue,citecolor=blue,pdftitle={Title},pdfauthor={Rubem Mondaini},pdfdisplaydoctitle=true,pdfduplex=DuplexFlipLongEdge]{hyperref}
\usepackage{amsmath,graphics,units}
\usepackage[table,xcdraw]{xcolor}
\usepackage{verbatim}
\usepackage[english]{babel}
\usepackage[T1]{fontenc}
\usepackage{color}
\usepackage{ulem}
\usepackage{physics}

\usepackage{braket}

\newcommand{\beq}{\begin{equation}} 
\newcommand{\eeq}{\end{equation}} 
\newcommand{\beqa}{\begin{eqnarray}} 
\newcommand{\eeqa}{\end{eqnarray}}

\begin{document}
\title{Topological Anderson  insulating phases in the interacting Haldane model}

\author{Jo\~ao S. Silva}
\email{jss.joaossilva@gmail.com}
\affiliation{Centro de F\'isica das Universidades do Minho e Porto, LaPMET,
Departamento de F\'isica e Astronomia, Faculdade de Ciencias,
Universidade do Porto, 4169-007 Porto, Portugal}

\author{ Eduardo V. Castro}
\email{evcastro@fc.up.pt}
\affiliation{Centro de F\'isica das Universidades do Minho e Porto,
Departamento de F\'isica e Astronomia, Faculdade de Ciencias,
Universidade do Porto, 4169-007 Porto, Portugal}

\author{Rubem Mondaini}
\email{rmondaini@csrc.ac.cn}
\affiliation{Beijing Computational Science Research Center, Beijing 100084,
China}

\author{Mar\'{\i}a A. H. Vozmediano}
\email{mahvozmediano@gmail.com}
\affiliation{Instituto de Ciencia de Materiales de Madrid, CSIC,
Sor Juana In\'es de la Cruz 3,  Cantoblanco,
E-28049 Madrid, Spain}

\author{M. Pilar L\'opez-Sancho}
\email{pilar@icmm.csic.es}
\affiliation{Instituto de Ciencia de Materiales de Madrid, CSIC,
Sor Juana In\'es de la Cruz 3,  Cantoblanco,
E-28049 Madrid, Spain}

\begin{abstract}
We analyze the influence of disorder and strong correlations on the topology of two-dimensional Chern insulators.  A mean-field calculation in the half-filled Haldane model with extended Hubbard interactions and Anderson disorder shows that the disorder favors topology in the interacting case and extends the topological phase to a larger region of the Hubbard parameters. In the absence of a staggered potential,  we find a novel disorder-driven topological phase with Chern number $C=1$, with the co-existence of topology with long-range spin and charge orders. More conventional topological Anderson insulating phases are also found in the presence of a finite staggered potential.
 
\end{abstract}

\maketitle

\section{Introduction} 
\label{sec_intro}

Topological phases of physical systems are one of the pillars of  modern condensed matter \cite{Topo2021}. The topological features of a material are established at the non-interacting level and the fate of topology in strongly correlated systems is a relevant topic of current research in the field~\cite{Rachel18}. 
Disorder, always present in real materials, also plays a vital role in the phase diagram of correlated electrons. Although strong disorder would be detrimental to topology -- eventually leading to trivial, Anderson localized phases in two-dimensional (2D) systems~\cite{Anderson58}--, disorder-induced topological phases (Anderson topological insulators)~\cite{AndTopo2009} are an exciting possibility proposed recently.
In this work, we explore the interplay of topology, disorder, and interactions using the Haldane model at half-filling \cite{H88} as a paradigm of topological Chern insulators in 2D. For that, we consider the extended Hubbard model with on-site and nearest neighbor (NN) $U$ and $V$ interactions, respectively, subjected to Anderson disorder $W$, and explore the ensuing mean-field phase diagram. 

\begin{figure}
\begin{centering}
  \includegraphics[width=0.95\columnwidth]{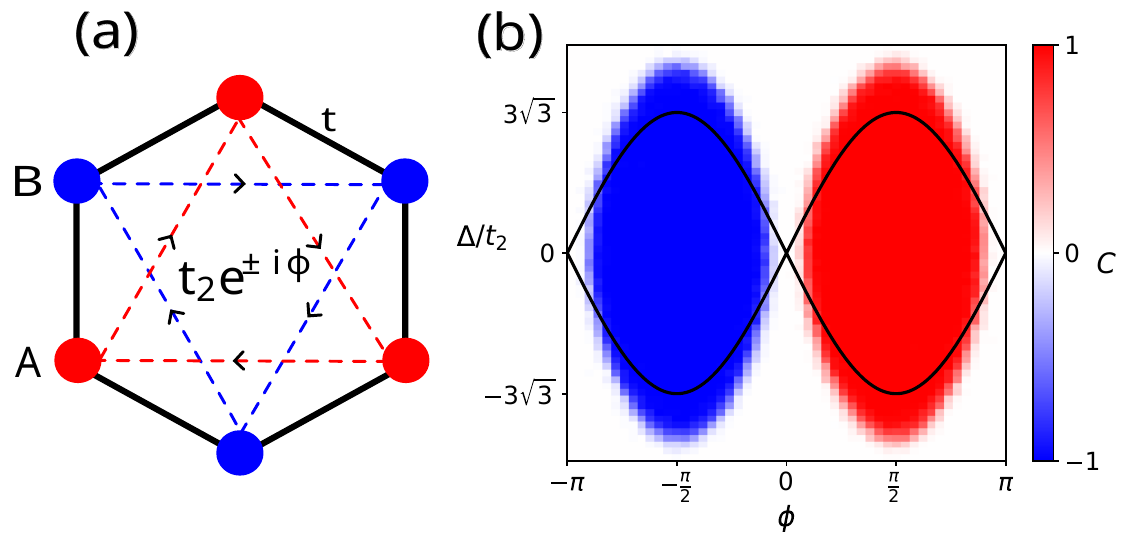}
  \par
\end{centering}
\caption{ (a) Schematic representation of the Haldane model on the honeycomb lattice with annotated terms for the hopping between nearest and next-nearest neighbors. (b) 
Phase diagram of the (non-interacting) spinless Haldane model as a function of the staggered potential and effective Haldane mass parametrized by the phase of the NNN hoppings. Here, the continuous lines give the clean case ($W=0$) result, whereas the colors map the topological regions in the presence of finite disorder ($W/t=4$). We have set $t_2 = 0.2t$. Linear lattice size used in (b) is $L=19$, and the Chern numbers are obtained by averaging over 100 disorder configurations. The Chern number is doubled in the spinful system.}
\label{fig1}
\end{figure}

\begin{figure*}[th!]
\begin{centering}
\includegraphics[width=0.99\textwidth]{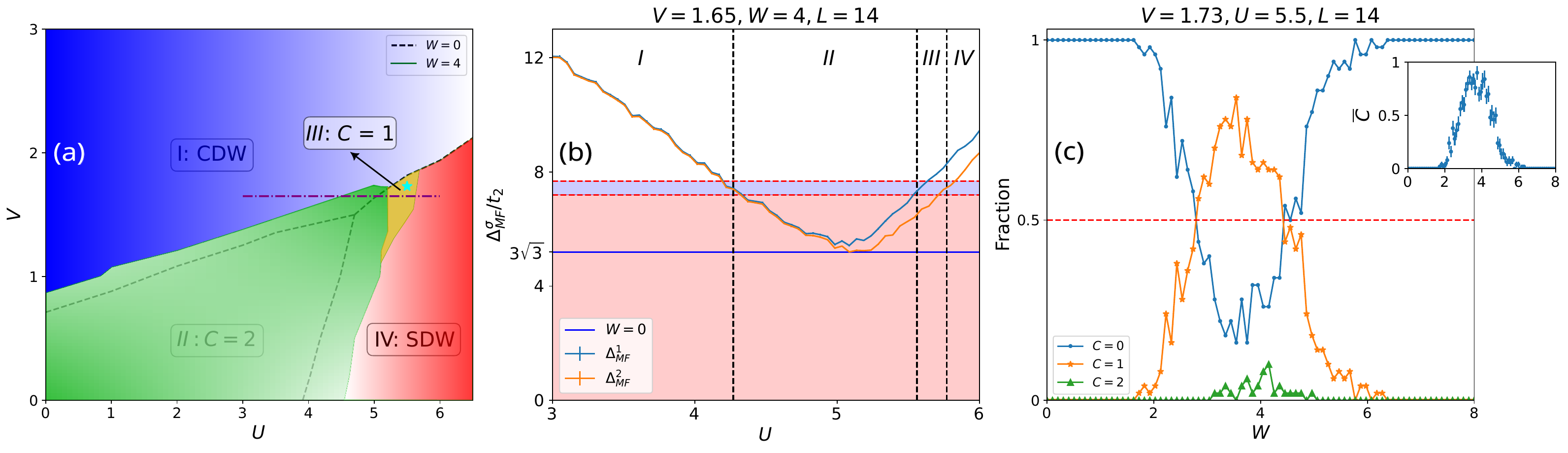} 
\par
\end{centering}
\caption{(a) Phase diagram of the extended Haldane-Hubbard model on the half-filled honeycomb lattice as a function of the interactions $U$ and $V$ with zero staggered potential. The dashed lines mark the different phases in the absence of disorder. Full lines separate the phases when Anderson disorder $W=4$ is included. Four different phases are observed, including I, a topologically trivial phase featuring charge order; II, a topological Chern insulator with total Chern number $C=2$; III, a topologically non-trivial region but with $C=1$ and IV, a region exhibiting spin-ordering that is topologically trivial. (b) Analysis of the effective staggered spin-dependent potential $\Delta_{\rm MF}^\sigma$ [see Eq.~\eqref{DeltaMF}] along a line cut in (a), see horizontal dash-dotted line, with $V = 1.65$ which explains the spontaneous symmetry breaking of SU(2) that leads to a $C=1$ phase: the red-shaded region maps the topologically non-trivial region for this disorder strength in the non-interacting regime, cf.~Fig.~\ref{fig1}(b). (c) The fraction of disorder realizations that result in a given Chern number $C$ for a specific point in the $C=1$ phase [star marker in (a)] as a function of the disorder strength $W$; the inset gives the corresponding average Chern number. Linear lattice size used in (b) and (c) is $L=14$.}
\label{fig2}
\end{figure*}

The Haldane model was originally set as a lattice model of spinless electrons on the honeycomb lattice with nearest neighbors ($t$) and complex next-nearest neighbors ($t_2 e^{{\rm i}\phi}$) hopping energy scales, as schematically shown in Fig.~\ref{fig1}(a). At half-filling, a sufficiently large staggered potential $\Delta$ ($|\Delta/t_2| > 3\sqrt{3}$) leads to a topologically trivial phase, whereas the value of $t_2/t$ (combined with its corresponding phase $\phi$) promotes the topological regime as shown by the continuous line in the phase diagram in Fig.~\ref{fig1}(b). These phases are identified by the value of the Chern number, which counts the number and chirality of edge modes via the bulk-boundary correspondence once open boundary conditions are considered.  In particular, with added disorder, these topological regions in the phase diagram are deformed, as indicated by colored regions in Fig.~\ref{fig1}(b), here for $W=4t$~\cite{haldane_quenched_dis}. In the presence of an added spin degree of freedom, the corresponding topological phases have a Chern number $C=\pm 2$.

As it is well known~\cite{Castro21} and will be further detailed later, an on-site interaction $U$ drives the system to a spin density wave (SDW) while the NN interaction $V$ promotes a charge density wave phase (CDW). Both are topologically trivial insulators with a finite local order parameter emerging for each order type. The phase diagram of the clean, interacting model in the mean-field approximation is shown in Fig.~\ref{fig2}(a) (dashed lines). We primarily aim to generalize these results to include quenched, uncorrelated disorder. An initial expectation is that a critical value of disorder strength will generally drive the topological insulator to a trivial Anderson insulator. Still, as we will see in what follows, the ensuing phases in the presence of interactions can be manifestly richer than that. 

The paper is organized as follows: In Sec.~\ref{sec_model_and_methods} we introduce the model and the methods we use to analyze its properties. Our main findings are presented in Sec.~\ref{sec_main_results}. We put these results in context comparing with previous works in the literature in Sec.~\ref{sec_literature}. In Sec.~\ref{ssec_c1}, we provide details on the nature of the new disorder-driven topological $C=1$ phase and discuss the effect of the disorder on the original phase boundaries of the clean phase diagram. The disordered phases arising with a finite staggered potential are reviewed in Sec.~\ref{ssec_trivialM}. Open questions and possible future works are discussed in Sec.~\ref{sec_future}. Technical details on the model and calculations can be found in Appendix~\ref{sec_model}.

\section{Model and Methods}
\label{sec_model_and_methods}
The Haldane model is described by the Hamiltonian~\cite{H88}
\begin{equation}
H_0  =  -t\displaystyle\sum_{\langle i,j\rangle}c_i^{\dagger}c_j
-t_2\sum_{\langle\langle i,j\rangle\rangle}e^{-{\rm i}\phi_{ij}}c_i^{\dagger}c_j
+  \Delta\displaystyle\sum_i \eta_i c_i^{\dagger}c_i,
\label{TBHmodel}
\end{equation}
where $c_i=A, B$ are defined in the two triangular sublattices that form the honeycomb lattice. 
The first term $t$ represents a standard
real nearest neighbor hopping that links the two triangular sublattices. The $t_2$ term represents a complex next nearest neighbor (NNN) hopping $t_2 e^{-{\rm i}\phi_{ij}}$ acting within each triangular sublattice with a phase $\phi_{ij}$ that has opposite signs 
$\phi_{ij}=\pm\phi$ (governing  the chirality) in the two sublattices.
The structure of the NNN hoppings is shown in Fig.~\ref{fig1}(a).
This term breaks time-reversal symmetry and opens a non-trivial topological gap at the Dirac points proportional to the magnitude of $t_2$.
We restrict the calculations to  $\phi=\pi/2$, since it maximizes the topological region in the non-interacting regime [see the full line in Fig.~\ref{fig1}(b)] and take the value $t_2 = 0.2t$. The last term in Eq.~\eqref{TBHmodel} represents a staggered potential ($\eta_i=\pm 1$). It breaks inversion symmetry and opens a trivial gap at the Dirac points, as seen in Fig.~\ref{fig1}(b). Spin doubles the degrees of freedom and  the Chern number is $C=\pm 2$ in the topological phases. 

The interacting Hamiltonian we consider has the form 
\begin{equation}
H_{\rm int}=U\sum_{i}n_{i,\uparrow}n_{i,\downarrow}+V\sum_{\langle i,j\rangle,\sigma,\sigma^{\prime}}n_{i,\sigma}n_{j,\sigma^{\prime}},
\label{eq_interactions}
\end{equation}
where $n_{i,\sigma}=c_{i,\sigma}^{\dagger}c_{i,\sigma}$ is the number operator. The on-site interaction term $U$ penalizes double occupancy, favoring, thus, a homogeneous charge distribution between the two sublattices. In this sense, it goes against the staggered on-site potential $\Delta$ and can favor topology to some extent. $U$ also
has the effect of polarizing the spin, and, over a critical value, it drives the system to a spin density wave insulator. The NN repulsive interaction $V$ favors sublattice charge imbalance and, similarly to $\Delta$,  goes against topology.

A chemical potential (Anderson) disorder is implemented by 
adding to the Hamiltonian the term $H_{\rm dis} = \sum_{i\in A, B}\varepsilon_{i}c_{i}^{\dagger}c_{i}$, with a uniform distribution of random local energies, $\varepsilon_{i}\in[-W/2, W/2]$. This on-site term will contribute to the mean field decoupling of the Hubbard $U$. Unless otherwise specified, the disorder averages were done using 50 disorder configurations.

A mean-field decoupling of Eq.~\eqref{eq_interactions} gives
\begin{equation}
\begin{aligned}&H_{\rm int}^{\rm MF}  =U\sum_{i,\sigma}\left[\langle n_{i,-\sigma}\rangle c_{i,\sigma}^{\dagger}c_{i,\sigma}-\langle c_{i,-\sigma}^{\dagger}c_{i,\sigma}\rangle c_{i,\sigma}^{\dagger}c_{i,-\sigma}\right]+\\
 & +V\left[\sum_{\langle i,j\rangle,\sigma,\sigma^{\prime}}\langle n_{j,\sigma^{\prime}}\rangle c_{i,\sigma}^{\dagger}c_{i,\sigma}-\sum_{\langle i,j\rangle,\sigma,\sigma^{\prime}}\langle c_{j,\sigma^{\prime}}^{\dagger}c_{i,\sigma}\rangle c_{i,\sigma}^{\dagger}c_{j,\sigma^{\prime}}\right]\ ,
\end{aligned}
\end{equation}
where the fields $\langle c_{i,\sigma}^{\dagger}c_{j,\sigma^{\prime}}\rangle$ are obtained self-consistently (see Appendix~\ref{sec_model}). Finally, the total calculated Hamiltonian reads $H = H_0 + H_{\rm int}^{\rm MF} + H_{\rm dis}$. 

From the total calculated Hamiltonian $H$ we define the  effective staggered spin-dependent potential $\Delta_{\rm MF}^\sigma$, 
\begin{equation}
\begin{aligned} \Delta_{\rm MF}^\sigma = \frac{1}{N} \sum_{i\in A} \frac{ \left| \xi_{i,\sigma} - \xi_{j_i,\sigma} \right| }{2},
\end{aligned}
\label{DeltaMF}
\end{equation}
where $j_i \in B$ is the NN of site of $i$ which belongs to the same unit cell, and $\xi_{i,\sigma}$ is a diagonal element of $H$
in the real space tight-binding basis $c^\dagger_{i,\sigma} |0\rangle = |i,\sigma \rangle$,
\begin{equation}
    \langle i, \sigma| H | i,\sigma \rangle \equiv \xi_{i,\sigma} = U\langle n_{i,-\sigma} \rangle + V \sum_{\Vec{\delta},\sigma^\prime}\langle n_{i+\Vec{\delta},\sigma^\prime} \rangle + \varepsilon_i .
    \label{xi}
\end{equation}
After disorder averaging, $\Delta_{\rm MF}^\sigma$ turns out to be an important quantity to understand the obtained results, as will be discussed in Sec.~\ref{sec_explain}.

\section{Main findings and antecedents} 

\subsection{Main results} 
\label{sec_main_results}

Figure~\ref{fig2} summarizes our main results. In particular, Fig.~\ref{fig2}(a) shows the phase diagram of the disordered, spinful Haldane model as a function of the extended Hubbard interactions $U$ and $V$ in units of the NN hopping parameter $t$. The Haldane parameters are chosen in the topological region of Fig.~\ref{fig1}(b) with zero staggered potential $\Delta=0$ and $\phi=\pi/2$. The dashed lines mark the different phases in the absence of disorder, for better comparison: the standard Chern insulator with Chern number $C=2$, and the $C=0$ SDW and CDW phases. Continuous lines separate the phases when Anderson disorder $W=4$ (in units of $t$) is included. The disorder is seen to enlarge the topological $C=2$ region and to generate a novel $C=1$ phase near the boundary of the three phases. This phase has long-range spin and charge orders. Figures~\ref{fig2}(b) and~\ref{fig2}(c) present an interpretation of this $C=1$, which which will be put forward in Sec~\ref{sec_explain}. 

Finally, we highlight that the experimental realization of the Haldane model~\cite{Jotzu_2014, HaldaneExp22} and the ability to realize strongly correlated Hubbard models using cold atom systems \cite{Greiner2002,Esslinger2010} give real prospects to emulate the spinful, extended Haldane-Hubbard model~\cite{RG18,Guardado-Sanchez2021}. With the ability to include disorder \cite{Schreiber2015,Choi2016}, the door is open to direct confirmation of the results of this work.

\subsection{Antecedents}
\label{sec_literature}

The effect of disorder and /or Hubbard interactions on the Haldane model has a long pre-history related with  the non-topological honeycomb lattice. The phase diagram in Fig.~\ref{fig2} substituting the 
CI phase with SM (semimetal) has been revisited over and over since the pioneer works \cite{Sorella92,muramatsuSL2010,sorellaNoSL2012,PhysRevX.3.031010}.
In this section we will only discuss the previous works in the literature that are closely related with our results.

\begin{itemize}
\item A $C=1$ phase in the clean, spinful Haldane model with only on-site Hubbard $U$ was found in \cite{HeFeng_2011,He_2012,Troyer_2016,Wang_2019,Roser19,Prok19,Yuanetal22} as an interplay of finite staggered potential  $\Delta$ and $U$. No $C=1$ was found in mean field calculations with $\Delta=0$. The new phase is spin-polarized and was termed "a topological spin density wave". An intuitive physical picture of this phase will be described in the next section. An important open question around this phase is whether or not it is an artifact of the used approximations, like mean field approximation,  since it was not found in a  dynamical cluster approximation in Ref.~\cite{Troyer_2016_2}. The $C=1$ phase was recently re-established with an exact diagonalization calculation in \cite{Yuanetal22}. Its stability against long range Coulomb interaction was examined in \cite{Prok19} using a diagrammatic Monte Carlo method.

\item Topological transitions in the extended Haldane--Hubbard model ($U$, $V$) with zero staggered potential and no disorder  ($\Delta=0$, $W=0$)  were studied in \cite{Castro21}. No $C=1$ phase was found there, except for a particular cluster used in the exact diagonalization and attributed to finite size effects. A variety of techniques led them to conclude  that topological and locally ordered phases do not coexist in the model. 

\item
Interestingly, a $C=1$ phase has also been found in the topological square lattice ($C=2$ in the non-interacting limit) \cite{Wang_2019} with $U$ and $V$ interactions and a sublattice potential   $\Delta=2$. A mean field calculation shows a $C=1$ phase named  (interaction-driven) antiferromagnetic Chern insulator (AFCI) by the authors.  As in previous works, this phase is not present when $\Delta=0$. 

\item The interplay of NN interaction $V$, disorder, and topology in the
spinless Haldane-Hubbard model was addressed in \cite{Castro2021}. A  topological Anderson insulator found in the non-interacting system with a finite staggered potential, was shown to be stable to the presence of sufficiently small interactions.

\end{itemize}

The study of the effect of disorder in the spinful extended Haldane--Hubbard model is clearly missing. Also missing from previous results is a $C=1$ phase with $\Delta=0$.

\section{Characterizing the new Anderson topological insulators}
\label{sec_explain}
\subsection{Phase diagram in the  $\Delta=0$ case.}
\label{ssec_c1}
\begin{figure}
\begin{centering}
\includegraphics[width=0.95\columnwidth]{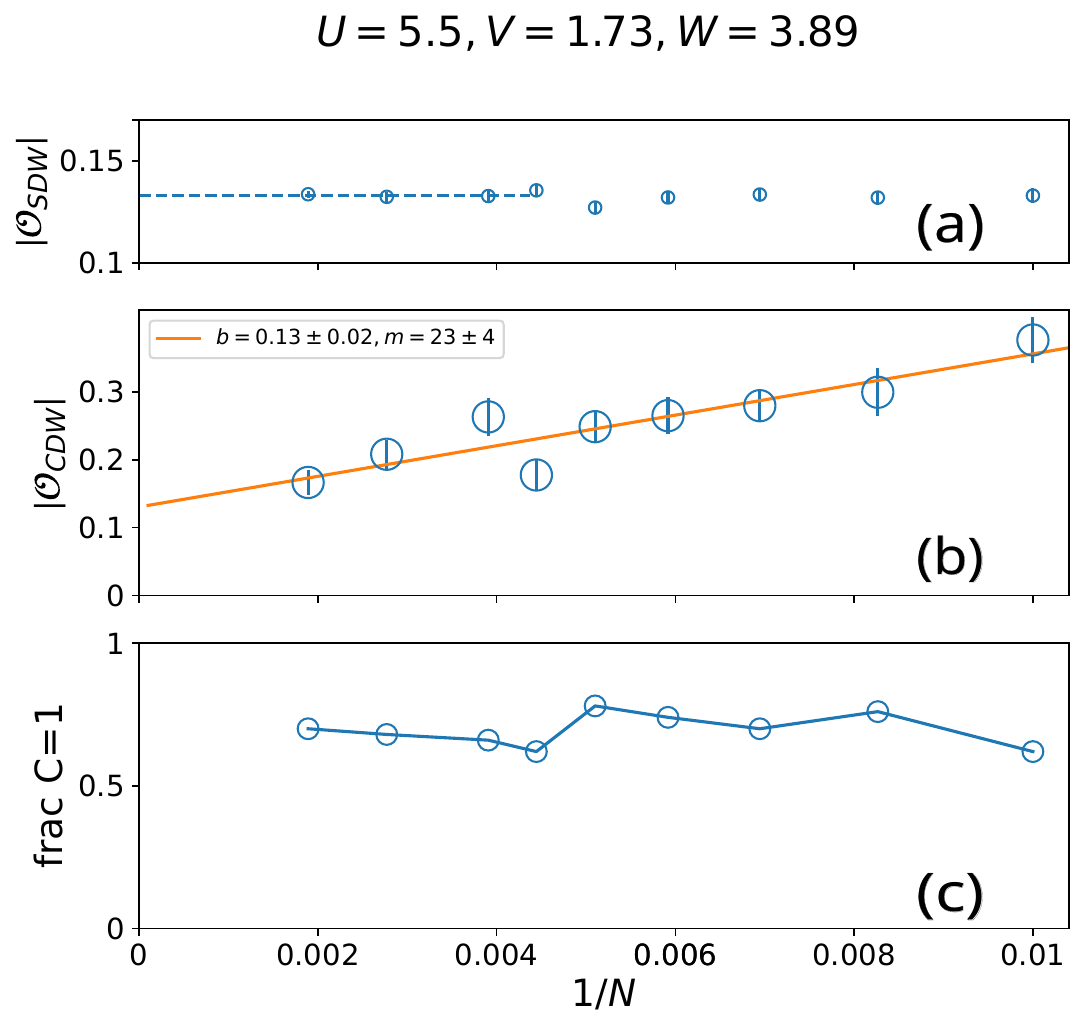} 
\par
\end{centering}
\caption{ Spin (a) and charge (b) order parameters in the $C=1$ phase, and fraction of disorder configurations which yield $C=1$ (c), as a function of the inverse of lattice size $1/N$, with $N=L^2$  the number of unit cells. The charge order parameter has been fitted to $\vert\mathcal{O}_{\text{CDW}}\vert=m(1/N) + b$. The error bars are the standard error of the mean (see Appendix~\ref{sec_model}).}
\label{fig3}
\end{figure}
\begin{figure}
\begin{centering}
\includegraphics[width=0.95\columnwidth]{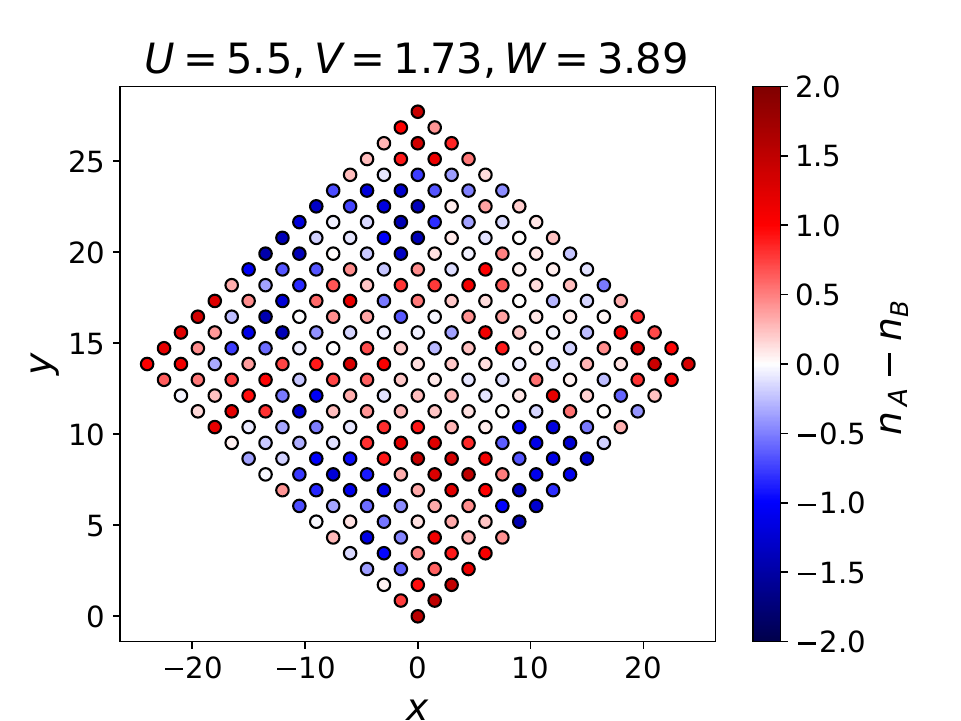} 
\par
\end{centering}
\caption{ Typical charge density imbalance in the $C=1$ phase for a given disorder configuration. Each circle represents the difference between the electron density of the two sublattices for a given  unit cell ($n_A - n_B$), where $n_\Gamma = \sum_{\sigma = \uparrow,\downarrow} \langle n_{i\in \Gamma,\sigma}\rangle $. Linear lattice size used is $L=17$.}
\label{fig4}
\end{figure}

The phase diagram of the $\Delta=0$ case is shown in Fig.~\ref{fig2}(a).
The most interesting finding there is 
the $C=1$ phase arising from the interplay of $U$, $V$, and $W$.  This is a topological Anderson insulator phase, highly disordered, showing a non zero spin polarization and charge inhomogeneities (electron-hole puddles) with a non zero mean value of the SDW and CDW order parameters. The  spin and charge order parameters are defined in Eq.~\eqref{eq_orderP}. Their evolution  as a function of the lattice size is shown in Fig.~\ref{fig3} ($N$ is the number of unit cells). The circles are calculated points with standard deviation of the mean as the error bars in  the vertical lines (see Appendix~\ref{sec_model}). As inferred from Fig.~\ref{fig3}(a), it is clear that the spin order parameter will remain finite in the thermodynamic limit. The CDW order parameter shows more oscillations but, as is evident from the fit in Fig.~\ref{fig3}(b) (see figure caption) it does not extrapolate to zero. To illustrate the robustness of the $C=1$ phase as the system size increases, we show in Fig.~\ref{fig3}(c) the 
fraction of $C=1$ disorder configurations as a function of $1/N$, where $N$ is the number of unit cells.  This result strongly indicates that the $C=1$ phase is not a finite-size effect. A typical configuration of the charge inhomogeneity in the $C=1$ phase is shown in Fig.~\ref{fig4} for $U=5.5$, $V=1.73$, and $W=3.89$.  This phase is at odds with the analysis in  Ref.~\cite{Castro21} where they found no co-existence of topological and long range ordered phases in the clean model. 
%


The $C=1$ phase described previously in the spinful Haldane model \cite{HeFeng_2011,He_2012,Zhu_2014,WuPeng_2015,Troyer_2016,Troyer_2016_2,Wang_2019,Yuanetal22} was due to  the interplay of a staggered potential and the local Hubbard $U$  interaction without NN interaction $V$ in the clean topological lattice. The $C=1$ phase was found in a narrow region between the two topologically trivial insulators induced by high values of the staggered potential (trivial insulator) and local $U$ interaction (Mott-Hubbard insulator). The exotic phase was dubbed a ``topological spin density wave'' and is of the same type as the one described here. 

\begin{figure}
\begin{centering}
\includegraphics[width=1.0\columnwidth]{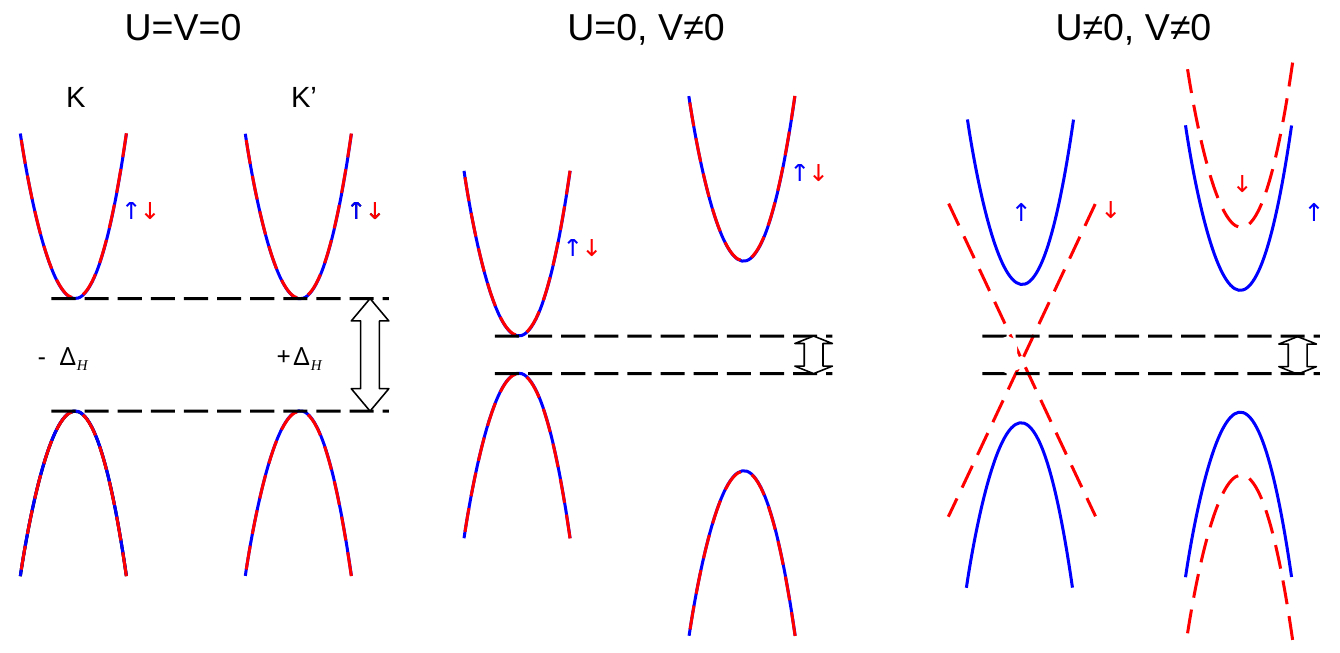} 
\par
\end{centering}
\caption{ Evolution of the Haldane bands around K and K' points of the Brilouin zone under the effect of  the interactions $U$ and $V$. $\Delta_H$ is the topological gap of the Haldane model.}
\label{fig5}
\end{figure}

An intuitive understanding of the $C=1$ phase works as follows: 
It is easy to see that, at the mean-field level, the CDW order parameter works like a staggered potential in the Haldane model, while the SDW order parameter works as a spin-dependent
staggered potential, with opposite sign for the two spin polarizations. The presence of both SDW and CDW order parameters will act as a trivial gap for one spin
polarization and reinforces the topological gap in the other. As a consequence, with increasing $U$, the bands for one spin polarization will become trivial while for the other they will still be topologically non-trivial. Since the Chern number is the sum of the two spin contributions, there will be a region
in parameter space where $C = 1$. 
This explanation of the $C=1$ phase is sketched in Fig.~\ref{fig5}. The left graph shows the bands of the Haldane model with zero staggered potential around the Dirac points K, K'. The bands are degenerated in spin and have an inverted gap. The CDW induced by a NN interaction $V$ splits the degeneracy of the valleys as shown in the middle panel. The SDW due to an on-site interaction $U$ lifts the spin degeneracy and moves the spin-polarized bands as indicated in the right hand panel. For a critical value of the parameters, the inverted gap closes in one of the spin-polarized bands that becomes topologically trivial giving rise to the $C=1$ phase. 
This explanation holds exactly for the $C=1$ phase observed with  $V=0$ and finite staggered potential  $\Delta$. A finite $\Delta$  explicitly breaks sublattice symmetry and leads to a finite charge imbalance equivalent to CDW.
%




Comparing with previous works in the literature one is tempted to think that the role played by $\Delta$ there, is taken by $V$ in our case.
However, the analysis of the clean extended Haldane--Hubbard model does not show the $C=1$ phase \cite{Castro21}. In the clean limit, the simultaneous presence of CDW and SDW is energetically unfavourable. It is disorder that, allowing the co-existence of topological and long range orders, permits the CDW order parameter to work as a trivial mass. This view is corroborated by Fig.~\ref{fig2}(b), where we plot the effective staggered spin-dependent potential $\Delta_{\rm MF}^\sigma$ given by Eq.~\eqref{DeltaMF}, for fixed $V$ and $W$ and varying  $U$ along the horizontal dash-dotted line indicated in the phase diagram of Fig.~\ref{fig2}(a). In the clean limit, this quantity plays the role of the effective gap for the spin-resolved bands shown in Fig.~\ref{fig5}. The first horizontal line in Fig.~\ref{fig2}(b) at $3\sqrt{3}t_2$ marks the topological transition in the clean, non-interacting Haldane model. In the presence of disorder, the topological region is wider [see Fig.~\ref{fig1}(b) at $\phi=\pi/2$], as signaled by the red-shaded region in Fig.~\ref{fig2}(b). The two horizontal dashed-red lines locate, within the uncertainty due to disorder averaging, the topological transition for the disordered Haldane model. It is seen that, for small $U$, the $\Delta_{\rm MF}^\sigma$ is spin-degenerate and have values above the topological transition. This agrees with the trivial CDW phase (region~I) in the phase diagram of Fig.~\ref{fig2}(a). Increasing $U$ leads to a decrease of $\Delta_{\rm MF}^\sigma$ which, for some critical interaction, falls below the topological transition line. This is compatible with the topological $C=2$ phase (region-II) in the phase diagram. Further increasing $U$, the spin-degeneracy of $\Delta_{\rm MF}^\sigma$ is lifted and the effective staggered potentials start to increase with $U$. At some point, one of the effective spin-dependent staggered potentials raises above the topological transition line while the other is still below. The system should then have $C=1$, fully compatible with region-III in the phase diagram. Astonishingly, the phase boundaries in Fig.~\ref{fig2}(b) are almost in quantitative agreement with the true phase diagram of Fig.~\ref{fig2}(a). This clearly indicates that, indeed,  $V$ is playing the role of $\Delta$ in the present case, as long as disorder is high enough.

The key role played by disorder is illustrated in Fig.~\ref{fig2}(c), where the fraction of disorder configurations for the three possible Chern values, $C=0,1,2$, is shown for a specific point in the phase diagram [blue star in Fig.~\ref{fig2}(a)] as a function of the disorder strength $W$. There is an optimal disorder for the $C=1$ fraction to dominate, and the average Chern number to approach $C=1$ (see inset). 
In the thermodynamic limit, it is expected that a finite region with $C=1$ develops around the optimal value of disorder.
As expected, for higher disorder the system is in a trivial phase. However, a trivial phase also shows up at small disorder, when only one of the ordered phases is established. We conjecture that the rather inhomogeneous CDW induced by disorder in the $C=1$ phase, as exemplified in Fig.~\ref{fig4}, may be the missing ingredient to stabilize the co--existence of SDW and CDW absent in the clean limit. Nevertheless, the fact that the $C=1$ phase only appears for high values of disorder indicates that it is a non-perturbative phase and that explanations based on perturbations around the clean limit have to be taken with caution.

Our results manifest the importance of disorder in the boundary regions close to phase transitions. 
We have analyzed the effect of the various parameters ($U$, $V$, $W$) on the boundaries between the $C=2$ phase and the CDW, SDW in Fig.~\ref{fig2}(a) for the $\Delta=0$ case. 
We have seen that Anderson topological insulating phases with $C=2$ can emerge beyond the phase transition lines of the clean model, as shown in Fig.~\ref{fig2}(a), in regions which were previously (for $W=0$) topologically trivial.
Over a critical (interaction dependent) value of disorder,  topology disappears and only trivial insulators remain.
Full phase diagrams for different sets of the parameters ($U$, $V$, $W$) can be found in Appendix~\ref{appOtherParams}.

\subsection{Disorder in the finite $\Delta$ phase diagram.}
\label{ssec_trivialM}
\begin{figure}[h]
\begin{centering}
\includegraphics[width=0.98\columnwidth]{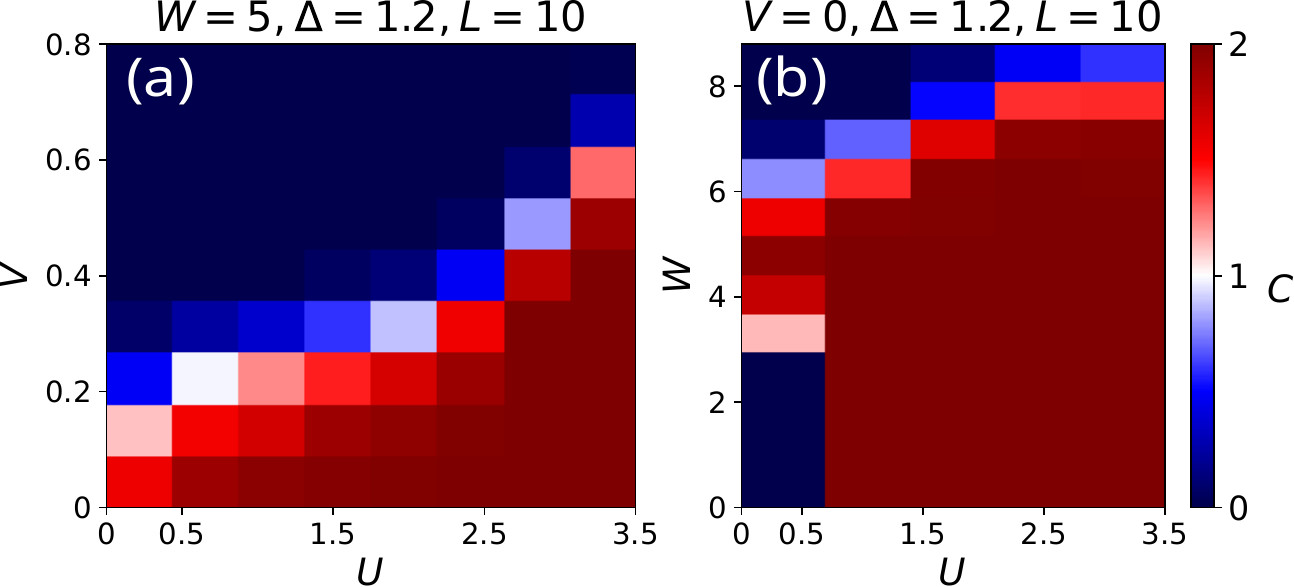} 
\par
\end{centering}
\caption{Phase diagram as a function of $U$ and $V$ for fixed $W$ (a), and as a function of $U$ and $W$ for fixed $V$ (b), for a finite $\Delta=1.2$. The colormap represents the value of the average Chern number. These were obtained for a system with linear size $L=10$ and averaging over 100 disorder configurations. The hopping parameters are the same as the ones used in previous results.}
\label{fig6}
\end{figure}

As mentioned before, an SU(2) broken $C=1$ phase  was previously found in the clean Haldane-Hubbard model  as a result of a competition of the SDW insulator driven by $U$ and the trivial insulator driven by the staggered potential $\Delta$ \cite{HeFeng_2011,He_2012,Zhu_2014,WuPeng_2015,Troyer_2016,Troyer_2016_2,Wang_2019,Yuanetal22}. 
 We have analyzed the influence of disorder and the interaction $V$ on that competition for a fixed value of $\Delta= 1.2$. 
 In Fig.~\ref{fig6}(a) we show the $(U,V)$ phase diagram for $W=5$. It can be seen that a $C=1$ phase appears [white pixel in Fig.~\ref{fig6}(a)] appears at much lower values of $U$ and $V$.
 The phase diagram in the $(U,W)$  plane is shown in Fig.~\ref{fig6}(b) for $V=0$. For small values of $U$, where the clean limit shows trivial behaviour, we see a re-entrant topological Anderson insulator phase with $C=2$ as disorder increases.
 In this context it is worth noticing the result discussed in~\cite{Troyer_2016} where it was seen that an explicit breakdown of SU(2)  by having different hopping amplitudes in the two sublattices led to the $C=1$ phase even at $U=0$.

It is worth noting that  plateau transitions $C=2\rightarrow1\rightarrow0$
are possible with increasing disorder at finite $\Delta$ and interactions.
 Plateau transitions $C=\pm2\rightarrow\pm1\rightarrow0$  with increasing disorder are conjectured to be ruled out in quantum Hall systems~\citep{hatsugai1999sum} and other Chern insulators derived from Dirac Hamiltonians~\citep{Song16}. In those systems, starting with $|C|\geq2$, a plateaus transition $\Delta C=\pm1$ is never observed with increasing disorder due to ensemble averaging over disorder realizations. Our results for finite disorder in the presence of interactions show that such a transition is possible, in particular if a finite trivial mass is also present.

\section{Open questions and future}
\label{sec_future}
As mentioned in Sec.~\ref{sec_literature}, an important open question  is to ascertain that the $C=1$ phase  is not an artifact of the mean field approximation \cite{Troyer_2016_2} or a finite size effect \cite{Castro21}. Exploring this region of the parameters with alternative methods as in \cite{Yuanetal22} will be very enlightening. 

Topological phase transitions between $C=1$ and $C=0$ or $C=2$ were found to be of  third order in the clean system \cite{HeFeng_2011,Shi2021}. Disorder makes the analysis of the nature of the phase transitions a hard problem that was left aside in this work, but studying  the nature of the phase transition between the $C=1$ and the surrounding phases is worth tackling in the future.  This is the problem of the phase transition between a standard and a topological Anderson insulator \cite{Chalker87} also related to the issue of localization  in quantum Hall systems \cite{Letal94,OAN07,09Murakami,Song16}. Another interesting issue is to analyze the structure of the topological edge states in the new phase and their evolution with increasing disorder. 

\begin{acknowledgments}
 JS and EC acknowledge financial support from FCT-Portugal through Grant No. UIDB/04650/2020. R.M.~acknowledges support from the NSFC Grants No.~NSAF-U2230402, No.~12111530010, No.~12222401, and No.~11974039. MPLS, MAHV and JS acknowledge the support of the Spanish Comunidad de Madrid grant S2018/NMT-4511 (NMT2D-CM). MAHV is also supported by the Spanish Ministerio de Ciencia e Innovaci\'on  grant PID2021-127240NB-I00.
This work was completed during a visit of MAHV to the Donostia International Physics Center (DIPC)
whose kind support is deeply appreciated.
\end{acknowledgments}

\appendix

\section{Mean-field calculations}
\label{sec_model}

In the main text, we describe the general methodology to extract the phase diagram of the model. Here, we introduce further technical details. In particular, the procedure to compute the mean field parameters $\langle c_{i,\sigma}^{\dagger}c_{j,\sigma^{\prime}}\rangle$ goes as follows: We initialize the parameters (or, in other words, we choose an initial condition for the system), we then
diagonalize the Hamiltonian and obtain its eigenvectors and energy spectrum, and re-calculate the mean field parameters,
\begin{equation}
\langle c_{i,\sigma}^{\dagger}c_{j,\sigma^{\prime}}\rangle=\sum_{E<E_{F}}\left[\psi_{i}^{\sigma}\left(E\right)\right]^{*}\psi_{j}^{\sigma^{\prime}}\left(E\right),
\end{equation}
with $E_{F}$ the Fermi energy and $\psi_{i}^{\sigma}\left(E\right)$
the wave function amplitudes. Finally, we define a convergence threshold
$\varepsilon$, and repeat the previous steps until $\left|\langle c_{i,\sigma}^{\dagger}c_{j,\sigma^{\prime}}\rangle_{I+1}-\langle c_{i,\sigma}^{\dagger}c_{j,\sigma^{\prime}}\rangle_{I}\right|<\varepsilon$,
with $I$ the iteration number.

Since a mean field method can be biased (i.e., the mean-field parameters reached after convergence can be heavily dependent on the choice of initial conditions), we employ a set of initial conditions, apply this procedure to each of them, and choose the solution that yields the lowest ground-state energy for this system (which can be calculated as the sum of the energies of the occupied eigenstates). The simple test of convergence we presented can, in some cases, prove to be very slow at reaching convergence. There are many ways of circumventing this issue; we chose to, after each iteration, define the new mean field parameters as the average of the previous two iterations. 

The topology of each phase is determined by the Chern number, which can be readily computed following  Fukui's method~\cite{FHS05}, modified to work with disordered systems~\cite{ZYetal13}, where translational
invariance is broken. 

The mean-field order parameters are obtained self-consistently for a fixed disorder configuration. Due to the finite size of the simulated clusters, we repeat the procedure for $N_{\rm dis}\sim 50-100$ disorder realizations. 
For the $C=1$ phase, we compute the Chern number for each disorder configuration. A perfectly quantized $C=1$ is obtained for over 60$\%$ of the disorder configurations in cluster sizes $N=10-23$, reaching 70$\%$ for the largest sizes [see Fig.~\ref{fig3}(c)].

To  characterize the onset and nature of various local orders, we define order parameters for charge density waves (CDW) and spin density waves (SDW) as
\begin{equation}
\begin{aligned}\mathcal{O}_{\rm CDW} & =\frac{1}{N}\sum_{\sigma}\left(\sum_{i\in A} \langle n_{i,\sigma} \rangle -\sum_{i\in B} \langle n_{i,\sigma} \rangle \right),\\
\mathcal{O}_{\rm SDW} & =\frac{1}{4N}\left(\sum_{i\in A}\left(\langle n_{i,\uparrow}\rangle- \langle n_{i,\downarrow}\rangle \right)-
\sum_{i\in B}\left(\langle n_{i,\uparrow}\rangle - \langle n_{i,\downarrow}\rangle \right) \right),
\end{aligned}
\label{eq_orderP}
\end{equation}
where $N$ is the number of unit cells. Since we are only interested in whether there is a net charge/spin imbalance, the sign of the structure factors is irrelevant, and we focus only on the absolute value of the order parameters.
 These quantities are calculated for each disorder configuration and subsequently averaged over all disorder configurations
as done with the Chern number. The standard error associated with the disorder averages (depicted, for example, as vertical bars on the data points of Fig.~\ref{fig3}) is calculated via the standard error of the mean of $N_{\rm dis}$ disorder realizations.

\section{Other sets of parameters}
\label{appOtherParams}

\begin{figure}[h]
\begin{centering}
\includegraphics[width=0.95\columnwidth]{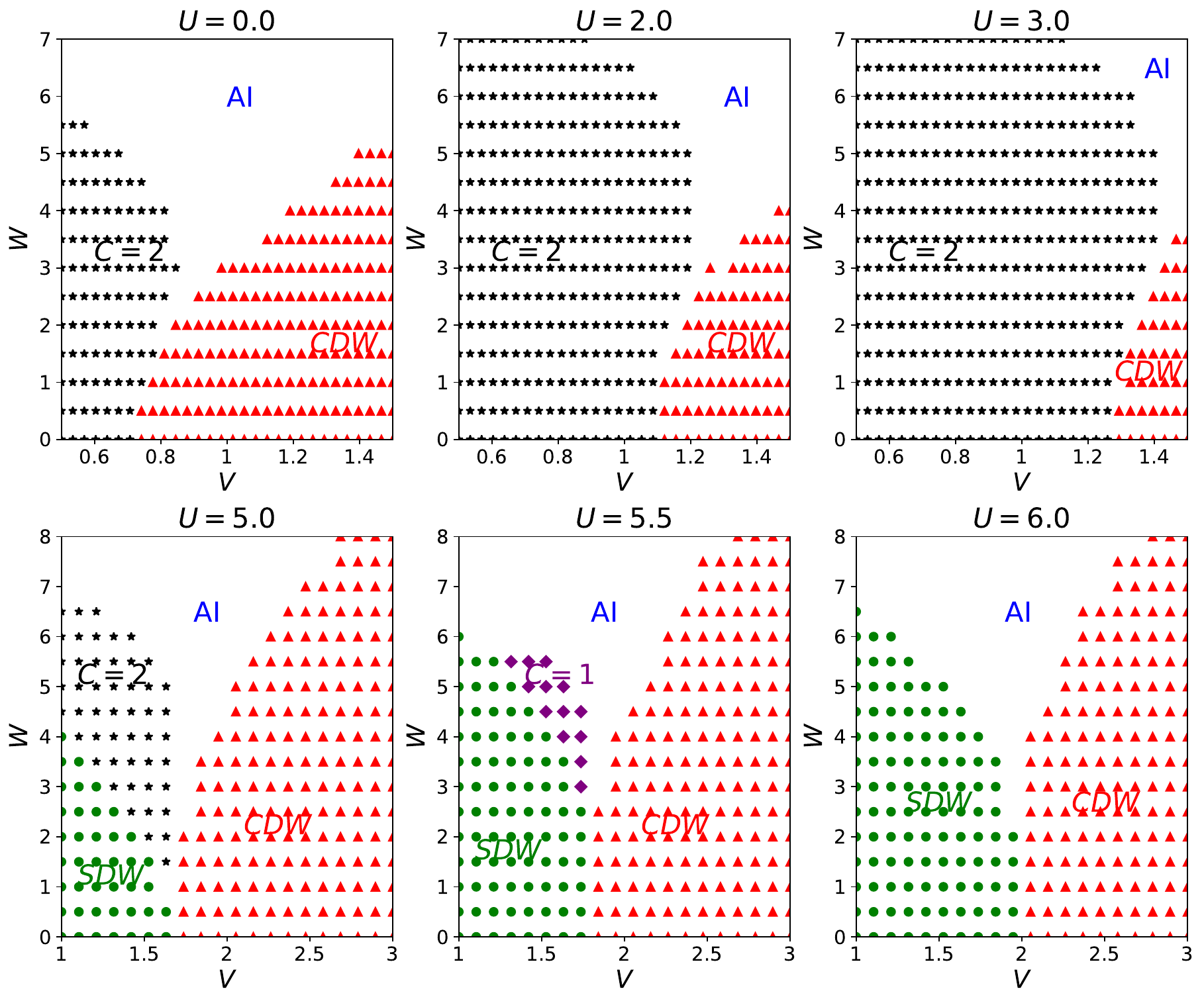} 
\par
\end{centering}
\caption{Phase diagrams as a function of $V$ and $W$ for different choices of $U$. These were obtained for a system with linear size $L=10$ and averaging over 50 disorder configurations. The hopping parameters are the same as the ones used in the results presented in the main text.}
\label{figapp1}
\end{figure}

\begin{figure}[h]
\begin{centering}
\includegraphics[width=0.95\columnwidth]{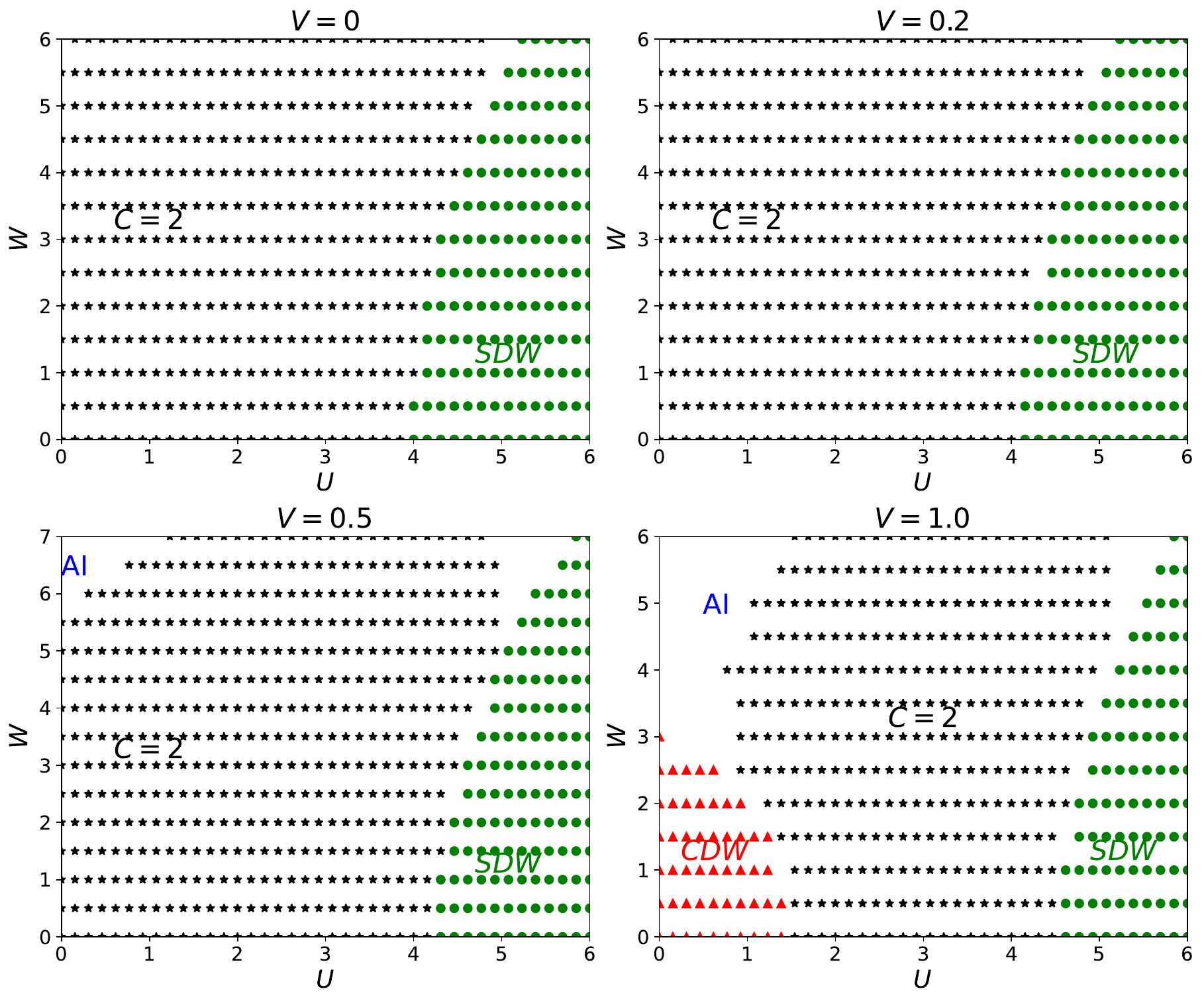} 
\par
\end{centering}
\caption{Phase diagrams as a function of $U$ and $W$ for varying choices of $V$. These were obtained for a system with linear size $L=10$ and averaging over 50 disorder configurations. The hopping parameters are the same as the ones used in  the results presented in the main text.}
\label{figapp2}
\end{figure}

Throughout section \ref{sec_explain}, we focused on the regions of the parameter space that corresponded to the generation of the $C=1$ phase with long-range ordering. In this Appendix we present a more systematic study of the parameter space for the proposed model.

The phase diagrams shown in Fig.~\ref{figapp1} for fixed $U$ shows that, as $U$ is increased (bellow a certain value), the topological region is enlarged, corroborating the idea that the on-site interaction term can favor topology to some extent. If $U$ is further increased (to values of between 5 and 6), the rich competition between the interaction and disorder terms leads to the emergence of the previously discussed $C=1$ phase. As expected, for stronger values of the interactions, the only phases that survive are trivial insulators (CDW and SDW for very large $V$ and $U$, respectively). For large disorder strengths, the long-range ordering breaks down, and only a trivial Anderson insulator phase survives. Similar conclusions are reached when analysing the phase diagram ($U$,$V$) for fixed NN interaction $V$ (see Fig.~\ref{figapp2}), with the addition that an increase in $V$ typically leads to a shrinking of the $C=2$ topological region.

\bibliography{Topo}

\end{document}